\documentclass{article}
\usepackage[top=0.85in,left=2.75in,footskip=0.75in]{geometry}
\usepackage{hyperref}
\usepackage{subcaption}
\usepackage{amsmath,amssymb}

\usepackage{changepage}

\usepackage[utf8x]{inputenc}

\usepackage{textcomp,marvosym}

\usepackage{cite}

\usepackage{nameref}

\usepackage{microtype}
\DisableLigatures[f]{encoding = *, family = * }

\usepackage[table]{xcolor}

\usepackage{array}

\newcolumntype{+}{!{\vrule width 2pt}}

\newlength\savedwidth

\raggedright
\usepackage[aboveskip=1pt,labelfont=bf,labelsep=period,justification=raggedright,singlelinecheck=off]{caption}

\makeatletter
\renewcommand{\@biblabel}[1]{\quad#1.}
\makeatother

\usepackage{lastpage,fancyhdr,graphicx,placeins}
\usepackage{epstopdf}

\usepackage[normalem]{ulem}

\fancyheadoffset[L]{2.25in}
\fancyfootoffset[L]{2.25in}
\newcommand{\E}{\mathbb{E}}
\newcommand{\Prob}{\mbox{Prob}}

\def\*#1{\boldsymbol{#1}}
\def\~#1{{\cal #1}}

\newcommand{\hx}{\hat{x}}

\begin{document}
\vspace*{0.2in}

\begin{flushleft}
{\Large
\textbf\newline{Simulation-free estimation of an individual-based SEIR model for evaluating nonpharmaceutical interventions with an application to COVID-19 in the District of Columbia}
}
\newline
\\
Daniel K. Sewell*\textsuperscript{1},
Aaron Miller \textsuperscript{2},
for the CDC MInD-Healthcare Program
\\
\bigskip
\textbf{1} Department of Biostatistics, University of Iowa, Iowa City, IA, USA
\\
\textbf{2} Department of Epidemiology, University of Iowa, Iowa City, IA, USA
\\
\bigskip

% Use the asterisk to denote corresponding authorship and provide email address in note below.
* daniel-sewell@uiowa.edu

\end{flushleft}
\section*{Abstract}
The ongoing COVID-19 pandemic has overwhelmingly demonstrated the need to accurately evaluate the effects of implementing new or altering existing nonpharmaceutical interventions.  Since these interventions applied at the societal level cannot be evaluated through traditional experimental means, public health officials and other decision makers must rely on statistical and mathematical epidemiological models. Nonpharmaceutical interventions are typically focused on contacts between members of a population, and yet most epidemiological models rely on homogeneous mixing which has repeatedly been shown to be an unrealistic representation of contact patterns.  An alternative approach is individual based models (IBMs), but these are often time intensive and computationally expensive to implement, requiring a high degree of expertise and computational resources.  More often, decision makers need to know the effects of potential public policy decisions in a very short time window using limited resources.  This paper presents a computation algorithm for an IBM designed to evaluate nonpharmaceutical interventions.  By utilizing recursive relationships, our method can quickly compute the expected epidemiological outcomes even for large populations based on any arbitrary contact network.  We utilize our methods to evaluate the effects of various mitigation measures in the District of Columbia, USA, at various times and to various degrees.  \verb!R! code for our method is provided in the supplementary material, thereby allowing others to utilize our approach for other regions.

%\linenumbers

\section*{Introduction}

In December 2019, severe acute respiratory syndrome coronavirus 2 (SARS-CoV-2), was discovered in Wuhan China \cite{zhu2020novel}. The virus causes the Coronavirus Disease 2019 (COVID-19), characterized by fever, cough, shortness of breath and other respiratory or flu-like symptoms. Severe cases can lead to pneumonia, respiratory failure, multi-organ dysfunction and death \cite{chen2020epidemiological,wang2020clinical,richardson2020presenting}. Since its discovery, the virus has rapidly moved across the globe and on March 11th, 2020 the World Health Organization declared COVID-19 to be a global pandemic. To date, over 27.6 million cases and over 900,000 deaths have been reported worldwide \cite{dong2020interactive}. Nearly all countries have identified COVID-19 cases, with 158 countries reporting greater than a thousand cases and 50 countries having recorded over a thousand deaths \cite{dong2020interactive}. As the size of the pandemic continues to grow experts expect COVID-19 will pose a significant threat for many months and potentially years. Thus, the burden not just to population health, but also the overall healthcare system, skilled and long-term care, and the global economy are likely to be substantial.  

Given the rapid growth with which COVID-19 has moved across the globe, policy makers have sought guidance to slow the spread, reduce the severity of the epidemic, or guide strategies for reopening. Consequently, many infectious disease models have been developed to forecast the trajectory of the current epidemic and to understand the likely impact of a range of interventions \cite{murray2020forecasting,anastassopoulou2020databased,neher2020covid19scenarios,weissman2020locally,SenForecasting2020,branas2020flattening,ImperialForecasting2020,LosAlamosForecasting2020,woody2020projections,MITForecasting2020}.
Models designed to capture certain aspects of the epidemic (e.g., forecasting mortality) may not be well suited for others (e.g., evaluating policy decision making), and so decision makers have increasingly had to navigate a range of diverse modeling approaches while attempting to find approaches that can meet the specific nature of a given setting \cite{reynoldsPressConf20200401}. In short, decision makers need information on the effects of public policy measures on an epidemic that is both timely and accurate.  From a modeler's perspective, this translates into a model which is both computationally efficient and which captures salient features of transmission through a population.

Many models have been developed to represent the disease transmission process and evaluate control measures related to COVID-19. Such models typically fall into one of two major categories: homogeneous mixing compartmental models and individual/agent-based models (IBMs). Equation-based models are typically less computationally expensive and can be implemented quicker than IBMs. For example, \cite{anastassopoulou2020databased} implemented an early forecast for COVID-19 in Hubei, China, using a mass action compartmental model; \cite{li2020substantial} developed a compartmental model with homogeneous mixing which was used to capture the undocumented cases of COVID-19; \cite{kucharski2020early} used a compartmental model with homogeneous mixing which allowed for time-varying reproduction number; and two web-based forecasting platforms applied an interactive deterministic compartmental model assuming homogeneous mixing and allow a user to alter a time-dependent intervention \cite{neher2020covid19scenarios,weissman2020locally}. Some approaches have tried to estimate the effect of social distancing measures on COVID-19 such as \cite{song2020epidemiological} which uses a stochastic mass action compartmental model and \cite{moghadas2020projecting} which used an age stratified mass action model. A limitation of equation-based compartmental models is that they rely on the assumption of homogeneous mixing, or mass action.

An alternative to equation-based compartmental models are IBMs, which provide a method for capturing heterogeneous mixing by simulating contacts among individual agents. A wide range of IBMs have been developed for COVID-19 \cite{shoukat2020projecting,ferguson2020report9,abueg_sep_2020,aleta_may_2020,braun_sep_2020,chang_may_2020,chao_apr_2020,cuevas_june_2020,dorazio_apr_2020,fang_feb_2020,gopalan_apr_2020,hoertel_jul_2020,kai_apr_2020,kerr_may_2020,koo_jun_2020,kucharski_apr_2020,rockett_jun_2020,shamil_jul_2020,silva_jul_2020,stutt_may_2020}. IBMs have been developed to evaluate numerous interventions including community lockdowns or closures \cite{chang_may_2020,chao_apr_2020,gopalan_apr_2020,hoertel_jul_2020,koo_jun_2020,kucharski_apr_2020,shamil_jul_2020,silva_jul_2020,ferguson2020report9}, contact tracing, isolation and quarantine \cite{abueg_sep_2020,aleta_may_2020,chang_may_2020,chao_apr_2020,ferguson2020report9}, mask wearing \cite{dorazio_apr_2020,hoertel_jul_2020,kai_apr_2020,silva_jul_2020,stutt_may_2020}, social distancing \cite{aleta_may_2020,braun_sep_2020,chang_may_2020,dorazio_apr_2020,ferguson2020report9,koo_jun_2020} and travel restrictions \cite{chang_may_2020}. IBMs have also been developed to simulate multiple contact settings (e.g., homes, schools, workplaces, public transportation, etc.)\cite{abueg_sep_2020,aleta_may_2020,chao_apr_2020,ferguson2020report9} and have been applied across a range of populations, cities and countries.\cite{aleta_may_2020,chang_may_2020,chao_apr_2020,ferguson2020report9,hoertel_jul_2020} However, not all IBMs accurately capture realistic contact networks, such as \cite{kerr_may_2020} which assumes that individuals' degrees follow a Poisson distribution.  Such a short-tailed degree distribution misses the profound impact that so-called superspreaders have on the outbreak\cite{stein2011superspreaders,hornbeck2012using}. \cite{brauer2017mathematical} states (p.120),
\begin{quotation}
[superspreaders] transmit infection to many other members of the population, while most infectives do not transmit infections at all or transmit infections to very few others. This suggests that homogeneous mixing at the beginning of an epidemic may not be a good approximation.
\end{quotation} These IBMs, however, come at a steep computational cost and can often necessitate a large time commitment as well as a high level of expertise to design and code efficiently.

IBM's are particularly well-suited to evaluate nonpharmaceutical interventions that focus on contacts between individuals. To accurately understand the effects of these types of interventions, we must more accurately model how individuals contact one another. This has been commented on by others; e.g., \cite{cui2019influence} wrote ``realistic mixing can be an important factor to consider in order for the models to provide a reliable assessment of intervention strategies'' (p.31). There is a large body of research showing how in many settings homogeneous mixing is inadequate for accurately modeling disease dynamics, such as \cite{bansal2007individual,del2013mathematical,feng2015elaboration,bioglio2016recalibrating,chowell2016mathematical,brauer2017mathematical,cui2019influence} (see \cite{del2013mathematical} and \cite{bioglio2016recalibrating} for further reference listings on this). As a recent concrete example, \cite{burghardt2016testing} evaluated model assumptions in the West African Ebola outbreak and stated, ``we see that alternative hypotheses for how [Ebola Virus Disease] spreads, such as homogeneous mixing and nearest neighbor interactions, provide quantitatively poorer agreement with data'' (p.3). While not a focus of this paper, one important aspect is preferential mixing based on age \cite{delvalle2007mixing}.  While this has been done successfully for age-stratified homogeneous mixing models (see, e.g., \cite{feng2020influence}), if a model is to incorporate age when evaluating interventions on contacts, we believe that it would be better to incorporate this into a realistic network model through social selection mechanisms.

One of the advantages of equation-based models over IBMs is the simplicity and speed with which the former can typically be implemented. \cite{rahmandad2008heterogeneity} The computational complexity of IBMs, especially when simulating large populations and complex contact networks, often require extensive programming, algorithmic efficiency, computing power and parallel processing. \cite{parker2011distributed,bisset2014indemics,leonenko2015using} Conversely, many equation-based models can be quickly implemented using more simplistic numerical computing environments or even basic spreadsheet software. \cite{jaffry2008agent} Thus, equation-based models may be utilized more quickly during an outbreak setting of an emerging infectious disease. Indeed, during the COVID-19 pandemic many of the earliest modelling estimates were produced by equation-based models \cite{lin2020conceptual,tang2020estimation,fang2020transmission,prem2020effect} or branching process models (that focused on early stages of an outbreak)\cite{hellewell_april_2020,kretzschmar_apr_2020,kucharski_apr_2020,peak_april_2017}. Some of the earliest examples of IBMs applied to COVID-19 involved the re-purposing of models designed for previous outbreaks (e.g., influenza or smallpox).\cite{chang_may_2020,chao_apr_2020,ferguson2020report9,kretzschmar_apr_2020} In the early stages of an outbreak, policy makers may need to quickly evaluate multiple types of complex interventions under a range of unknown parameter values, and some policy actions may more well-suited to IBMs.\cite{currie2020simulation} Thus, there is a need to develop methods that can produce estimates consistent with IBM-based approaches while offering the computational efficiency typically represented by most equation-based models.

The contribution of this paper is primarily to present a computationally efficient estimation method for a network-based IBM which can be used for evaluating nonpharmaceutical interventions such as implementing or lifting social distancing measures or implementing universal personal protective equipment (PPE).  Our second contribution is to use this method to provide critical information on the effects of various nonpharmaceutical mitigation measures in the District of Columbia (D.C.) at various times and to various degrees.

\section*{Methods}
\label{sec:methods}
\subsection*{Overview}
\label{sec:methods-subsec:overview}
Our proposed method for modeling disease transmission dynamics through a susceptible-exposed-infectious-recovered (SEIR) model relies on a contact network rather than mass action assumptions. It focuses on the individuals in the population, but rather than simulating disease over an individual-based model and averaging the results to obtain an estimated epidemic curve, our method directly estimates the probability that a particular individual is infectious at a particular time.  Through the use of recursive relationships, the expected number of infected individuals can be efficiently computed at each time point.

The primary benefit of our proposed approach is the ability to obtain the expected epicurve from an SEIR IBM analytically, thereby saving practitioners from having to run thousands of simulations for each possible parameter configuration and counterfactual scenario. Other strengths of the method include the following.  First, it relies on a realistic or observed contact network.  This is in contrast to assumptions of homogeneous mixing, or that the contact degree distribution is not heavy tailed (e.g., Poisson).  This crucially allows us to capture the effects of superspreaders\cite{stein2011superspreaders,hornbeck2012using}, burstiness of the epidemic (i.e., the tendency for an epidemic to show alternating patterns of slow and rapid growth, as opposed to a steady growth curve)\cite{karsai2011small,colman2015memory}, and other salient features of realistic contact graphs. Second, our approach explicitly captures the way nonpharmaceutical interventions can affect the disease transmission through quarantining/social distancing or reducing the risk that a susceptible-infective contact will lead to a new transmission through, e.g., personal protective equipment (PPE).

Our method is limited in that some disease characteristics are simplified in exchange for more accurate contact patterns and computational efficiency. Specifically, it assumes that everyone who becomes infected experiences a constant latent period and is able to transmit the disease to their set of regular contacts (either directly or through the environment) for the same amount of time.  While these quantities can be estimated from the data, it does not reflect the varying lengths of time individuals are susceptible nor the varying lengths of time individuals are infectious before recovering, dying, or being effectively isolated. In addition, it may be that the population is partitioned such that subpopulations have varying levels of infectivity and susceptibility; for example, susceptible individuals of different ages may be more or less easily infected by infectives. While our approach allows (1) contacts patterns to differ between subpopulations and (2) transmission rates to vary over time, we have not accounted for individual level or dyadic level transmission rates.

While our proposed IBM may not be the most realistic model, as IBMs can always be more nuanced, it offers a significant computational benefit. Our proposed approach can be used to compute the expected results from a network-based IBM with large populations using limited computing resources, and our method is easily and quickly deployed (a small \texttt{R} package for implementing this methodology is provided in the supplementary material). This computational efficiency allows a user to quickly explore the effects of public policy changes on social distancing or universal PPE interventions, thereby providing decision makers a timely method of evaluating, for example, when and to what degree social distancing measures should be implemented or relaxed.

\subsection*{Approach}
\label{sec:methods-subsec:approach}
We begin by presenting the setup and notation we will use.  For a population of size $N$, let the initial probability that an individual is infected be denoted as $p_0$.  For $j=1,2,\ldots,N$, let $x_{tj}=1$ if individual $j$ is {\it infective} at time $t$, $t=1,2,\ldots,T$, and 0 otherwise.  The probability that $x_{tj}=1$, $\E(x_{tj})$, is denoted by $\hat x_{tj}$. The $N$ individuals in the population are connected through a contact graph which is represented by a $N\times N$ adjacency matrix $A$ such that $A_{ij} =1$ if $i$ and $j$ can contact each other and 0 otherwise. The probability that a susceptible individual $j$ is infected by an infective neighbor on the contact graph at time $t$ is denoted by $p_t$.  If this event occurs, the susceptible enters a latent period of $D_E$ days where they have been exposed but are not yet infectious. This latent period is immediately followed by a period of $D_I$ days where the individual is infective, and hence $\sum_t x_{tj} = D_I$ $\forall j$.  Let $\~I_t$ denote the set of infectives at time $t$, and let $\~N_j$ denote the neighbors of $j$ on the contact graph.  At time $t$, let $Q_{tj}$ denote the event that susceptible individual $j$ is self-quarantining, and let this event occur with probability $q_{tj}$.  Let $\iota$ denote the daily probability of a susceptible individual importing the disease from outside the population of interest.  Finally, let $H_{tj}$ denote the event that $j$ is successfully infected at time $t$ (and hence will enter the latent period for $D_E$ days), where $H_{0j}$ corresponds to the event that $j$ is the outbreak's initializer/patient zero. We denote the probability that an event, such as $H_{tj}$, has occurred by $\Prob(H_{tj})$, and the probability that it has not occurred by $\Prob(H_{tj}')$, and for conditional probabilities we use $|$ to separate the unknown event (left) from the events which we know to have occurred (right).

The goal of the analysis is to estimate the expected number of new infections each day- or equivalently each day's expected cumulative number of infections- according to the individual-based model described above.  To achieve this, we first focus on estimating the probability of being infective for each individual in the population on each day, i.e., $\hat x_{tj}$.

For the first few days of the outbreak, the only infective(s) will be the outbreak initializer(s).  If the latent period is longer than the infectious period then there will be one or more days with zero infectives.  If the opposite is true ($D_E < D_I$), then following these first few days there will be a period where the probability an individual is infective equals the probability that they are either the outbreak initializer or were infected within the first $t-D_E$ days.  After this, the probability that an individual is infective equals the probability that they were not an initializer and were infected within a moving window such that they have passed the latent period but have not yet recovered. To put this concretely in mathematical terms, we have the following.  For $1\leq t \leq \min(D_E,D_I)$, 
\begin{equation}
     \hx_{tj} = p_0.
     \label{eq:x1j}
\end{equation}
If $D_E > D_I$, then for $D_I < t\leq D_E$,
\begin{equation}
    \hx_{tj} = 0,
\end{equation}
else if $D_E < D_I$, then for $D_E < t \leq D_I$,
\begin{equation}
    \hx_{tj} =  \Prob\left( \bigcup_{s=0}^{t-D_E}H_{sj} \right).
\end{equation}
Finally, we have for $ t > \max(D_I,D_E)$,
\begin{equation}
    \hat x_{tj} = \Prob\left(\left\{\bigcap_{s=0}^{\max(0,t - D_E - D_I)}H_{sj}'\right\} \cap \left\{ \bigcup_{s=\max(1,t-D_E-D_i+1)}^{t-D_E} H_{sj} \right\}  \right).
\end{equation}

If individual $j$ is still susceptible at time $t$, they can become infected by either importing the disease from outside of the study population or by being infected by an infective neighbor on the contact graph. This latter method requires both that $j$ is not quarantined at time $t$ and that at least one infective neighbor infects $j$.  Hence by the law of total expectation we have that
\begin{align*}
    \Prob\left(H_{tj} \Big| \bigcap_{s<t}H_{sj}'\right) & = \iota + (1-\iota)(1-q_{tj})\Big( 1 - \E(\Prob(\mbox{no infectives infect } j|Q_{sj}',\*x_t) \Big) &\\
    & = \iota + (1-\iota)(1-q_{tj})\left( 1 - \E\left( \prod_{i\in\~N_j\cap\~I_t}(1-p_t) \right) \right) &\\
    & = \iota + (1-\iota)(1-q_{tj})\left( 1 - \E\left((1-p_t)^{\*x_t'A_{\cdot j}} \right) \right), &
\end{align*}
where $\*x_t = (x_{t1},x_{t2},\ldots,x_{tN})'$ and $A_{\cdot j}$ is the $j^{th}$ column of the adjacency matrix $A$.  Using a first order Taylor's expansion around $\hat{\*x}_t$, we have
\begin{align} \nonumber
    &\Prob\left(H_{tj} \Big| \bigcap_{s<t}H_{sj}'\right) & \\ \nonumber
    & \approx \iota + (1-\iota)(1-q_{tj})\left( 1 - \E\left((1-p)^{\hat{\*x}_t'A_{\cdot j}} + (\*x_t - \hat{\*x}_t)'\nabla \left.\left\{ (1-p_t)^{\*x_t'A_{\cdot j}} \right\}\right|_{\*x_t = \hat{\*x}_t} \right) \right), & \\  
    & =  \iota + (1-\iota)(1-q_{tj})\left( 1 - (1-p)^{\hat{\*x}_t'A_{\cdot j}}  \right), &  
    \label{eq:probH_it1}
\end{align}
or equivalently
\begin{align}
    \Prob\left(H_{tj}' \Big| \bigcap_{s<t}H_{sj}'\right) & \approx 
    (1-\iota)\left(q_t + (1-q_t)(1-p_t)^{\hat{\*x}_t'A_{\cdot j}}\right).& 
    \label{eq:probH_it2}
\end{align}
For ease of notation, we let this quantity in (\ref{eq:probH_it2}) be notated as $f_{tj}$.

Combining Eq (\ref{eq:x1j})-(\ref{eq:probH_it2}) yields the following.
\begin{align} \nonumber
    1<t\leq \min(D_E,D_I),\hspace{1pc} \hat x_{tj} & = p_0, &\\ \nonumber
    \min(D_E,D_I)< t \leq \max(D_E,D_I),\hspace{1pc} \hat x_{tj} & = \begin{cases}
    1 - (1-p_0)\prod_{s=1}^{t-D_E}f_{sj} & \mbox{ if } D_E < D_I \\
    0 &\mbox{ if } D_E > D_I
    \end{cases} &\\
    t > \max(D_E,D_I) \hspace{1pc} \hx_{tj} & = \alpha_{\max(0,t - D_E - D_I)j}\left( 1 - \prod_{s=\max(1,t-D_E-D_i+1)}^{t-D_E} f_{sj}\right).
\end{align}
Here $\alpha_{tj}$ is defined to be the probability that individual $j$ is still susceptible by time $t$.  These quantities can be computed recursively in the following manner.  First, $\alpha_{0j} := \Prob(H_{0j}') = 1-p_0$.  Subsequently for $t>0$ we have 
\begin{align} 
	\alpha_{tj} &:= \Prob\left(\bigcap_{s=0}^tH_{sj}'\right) = \alpha_{(t-1)j}f_{tj}.
\end{align}

These recursions then allow us to compute the quantity of interest, namely the number of infections we expect to have by time $t$.  This is derived from the expected number of susceptible individuals:
\begin{align} \nonumber
\E(\#\mbox{ infected}) & = \mbox{Pop. size } - \E(\#\mbox{ susceptible})& \\
&= N - \sum_{j=1}^N \alpha_{tj}.
\end{align}

The parameters to be estimated include the length of the disease's latent period ($D_E$), the number of days an individual is infective ($D_I$), the probabilities that a contact leads to a transmission event ($p_t$), and the probabilities that an individual is quarantined at time $t$ ($q_t$).  The initial probability $p_0$ may also need to be estimated, but in many cases this will be a known (small) number, such as $1/N$ for a single initializer.  Given a contact graph, prior to social distancing and any intervention which may affect transmission probabilities the quantities $p_t$ may be related to the more intuitive reproduction number, or $R_0$, defined to be the expected number of new infections generated by a given infective.  This can be computed as 
\[
R_0 = 
(\mbox{\# days infectious})
\times \E(\mbox{\# contacts per day})
\times \Prob(\mbox{S-I contact leads to new transmission}),
\]
and hence
\begin{equation}
    p_t = \frac{R_0}{D_I\bar d},
\end{equation}
where $\bar d$ is the average degree in the contact graph $A$, i.e., the average number of neighbors in $A$.  Estimation can be performed via ordinary least squares (OLS), i.e., minimizing the sum of squared differences between the observed and the expected number of daily cases.

\section*{COVID-19 in District of Columbia}
\subsection*{Case Data}
We used publicly available data collected by The COVID Tracking Project \cite{covidtracking} for Washington D.C.. This data source provides the daily number of lab confirmed cases which we used to train our IBM. We used data from the first case at the beginning of March until the end of May when protests over the killing of George Floyd led to a large number of unexpected potential transmission events.

To obtain estimates of the number of infections, we used results from a study involving a random sample of individuals to test in Indiana \cite{menachemi2020population} which estimated the infection fatality rate (IFR) to be 0.58\%.  We then used a 30-day rolling window to compute the reporting rate for each day; deaths were shifted by 16 days, as this was the average time between symptom onset and death \cite{sanche2020high}.  That is, for day $t$, we looked at the reporting rate $RR_t$
\[
RR_t = \frac{\sum_{s=t-29}^t \mbox{\# reported cases on day $s$}}{\left(\sum_{s=t-13}^{t+16} \mbox{\# reported deaths on day $s$}\right)/IFR}
\]
These daily reporting rates were then used to rescale the number of reported cases to obtain an estimate of the daily number of infections.

\subsection*{Contact graph}
We based our contact graph on a study of inter-personal daily contacts in Hong Kong. \cite{kwok2018temporal} In this study, 1,450 individuals were recruited from 857 households, where a contact was defined to be a social encounter which included a ``face-to-face conversation or touch (such as handshake, a kiss, games and sports or similar events involving body touch)'' \cite{kwok2018temporal}.  As was consistent with other studies, they found the distribution of contacts to be heavy tailed, which has profound effects in the context of infectious disease as this indicates potential super-spreaders.  Overall, individuals had on average 12.5 contacts per day.  This matches closely with an ecological momentary assessment study of individuals in upstate New York which found the average number of daily contacts to be 12\cite{zhaoyang2018age}.  We then constructed a network of 705,749 nodes by randomly assigning each node a number of stubs drawn from a negative binomial distribution with mean 12.5 and dispersion parameter 1.3.  Each stub was randomly connected to another node's stub, and in this way we preserved the desired degree distribution.

\subsection*{Mobility data}
We estimated individual response to quarantine policy measures and the corresponding mobility patterns using Google Community Mobility Reports \cite{aktay2020google}. The authors complied with the terms and conditions of \small \url{https://www.google.com/covid19/mobility/}. \normalsize Google Community Mobility Reports use geolocated mobile phone data to assess various populations' movement patterns.  This was measured by first capturing a baseline rate of movement and then evaluating deviations from this baseline rate.  These baseline measurements were taken from the period 3 January, 2020 - 6, February, 2020, and were broken down by day of the week and by category (workplace, retail and recreation, etc.).  We chose to focus on the category of retail and recreation, as this seemed to best reflect voluntary movement patterns.  We used ordinary least squares (OLS) to find a parameter by which we could multiply the  deviations given in the Google Community Mobility Reports in order to obtain a daily estimate of the probability an individual would self-quarantine.

\subsection*{Scenarios}
\begin{figure}
    \centering
    \includegraphics[width=0.7\textwidth]{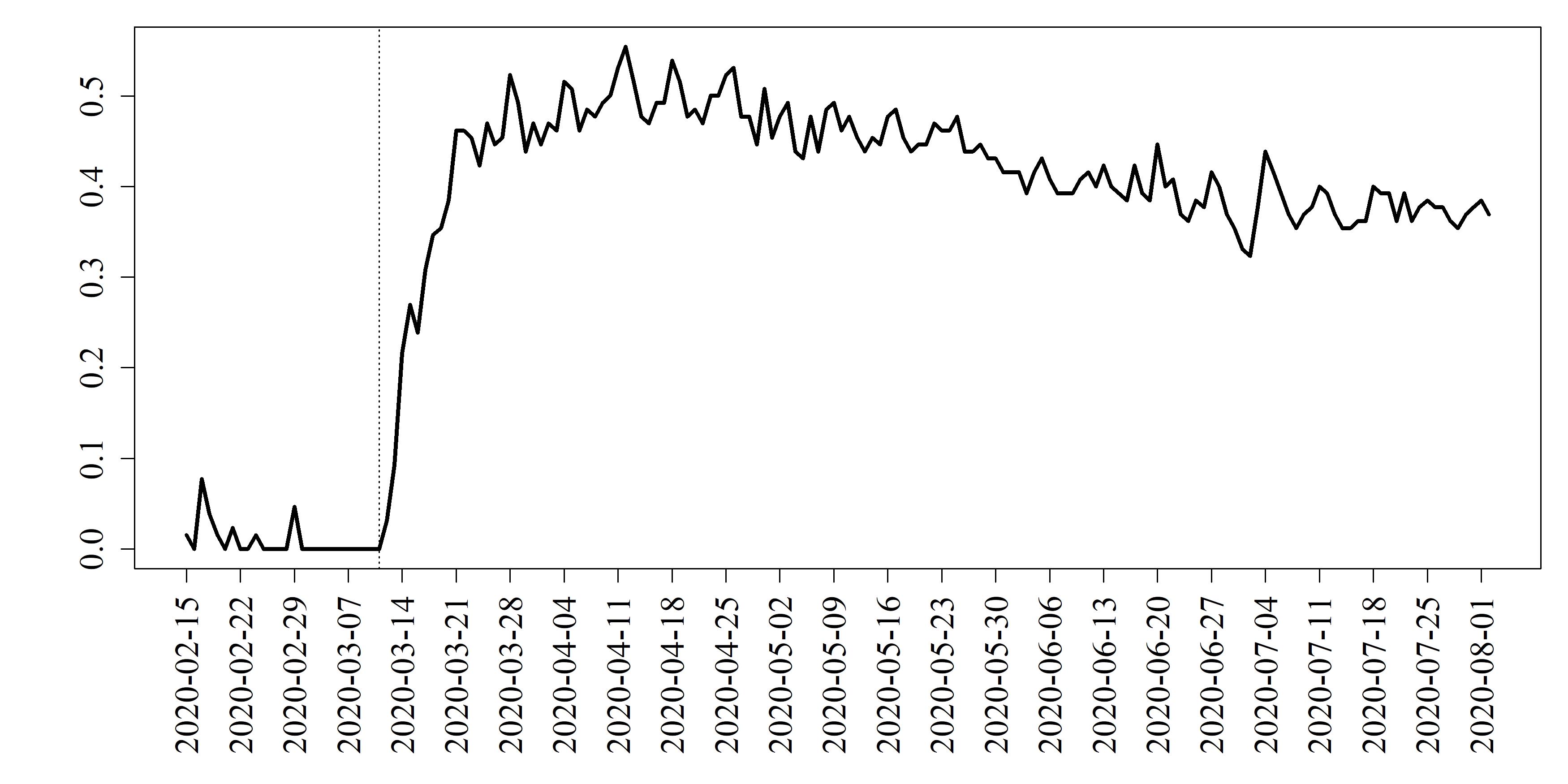}
    \caption{{\bf Estimated quarantining levels.}\\
    Estimated levels of quarantining based on the Google Community Mobility Reports.  The values in these reports have been scaled by a quantity estimated by training our IBM. The vertical line corresponds to the Declaration of Emergency issued by the mayor of D.C.}
    \label{fig:mobility}
\end{figure}

On March 11, 2020 the mayor of D.C. issued a State of Emergency declaration  \cite{raifman2020covid19}.  From the mobility data it is clear that this prompted a sudden shift towards self-isolation (see Figure \ref{fig:mobility}).  In addition, a face mask mandate was issued on April 15 \cite{raifman2020covid19}.  We trained our model using OLS on the daily number of infections, estimating the effect of the lockdown as well as the effect of the face mask mandate; the latter of these was estimated as a rescaling of the transmission parameter. 

We then computed the expected number of daily infections were the declaration of emergency to have occurred one week earlier and one week later than March 11; this was done by shifting the mobility data either backward or forward a week.  We also considered the scenario where there was no declaration of emergency.  We then evaluated the effect of replacing the lockdown via the declaration of emergency with a face mask mandate for the dates March 4, 11, and 18.  We additionally evaluated the effect of supplementing the lockdown with a face mask mandate, again for the same three dates.  Since the dates under consideration were so early in the course of the epidemic, we assumed that it would take some time before the same level of mask-wearing compliance as seen on April 15 would be observed.  We therefore allowed the transmission rate to change smoothly from pre-mandate to post-mandate levels through a Gaussian kernel such that after ten days the transmission rate was 90\% through the transition towards its final value.

We further estimated the effects of reimplementing social distancing for only a strategic portion of the population.  This was determined by implementing social distancing for those individuals with the highest number of contacts (e.g., healthcare professionals rotating between long term care facilities) while maintaining no social distancing measures for the remainder of the population. We evaluated the effects of a declaration of emergency on March 11 affecting only the top 10\% as measured by number of contacts and for the top 50\% both with and without an accompanying face mask mandate. In both cases, those designated to self-quarantine did so at the rate observed from the mobility data (see Figure \ref{fig:mobility}), while those not designated did not quarantine.

Lastly, we estimated the effect of delaying the lockdown induced by the declaration of emergency by 1, 2, and 3 weeks while implementing a mask mandate.  For all lockdown delays, this mask mandate was implemented on March 11 (the day of the observed declaration of emergency), and as before we allowed a gradual rollout of the mask mandate.

\subsection*{Results}
\begin{figure}
    \centering
    \captionsetup[subfigure]{justification=centering}
    \begin{subfigure}[a]{\textwidth}
    \includegraphics[width=0.7\textwidth]{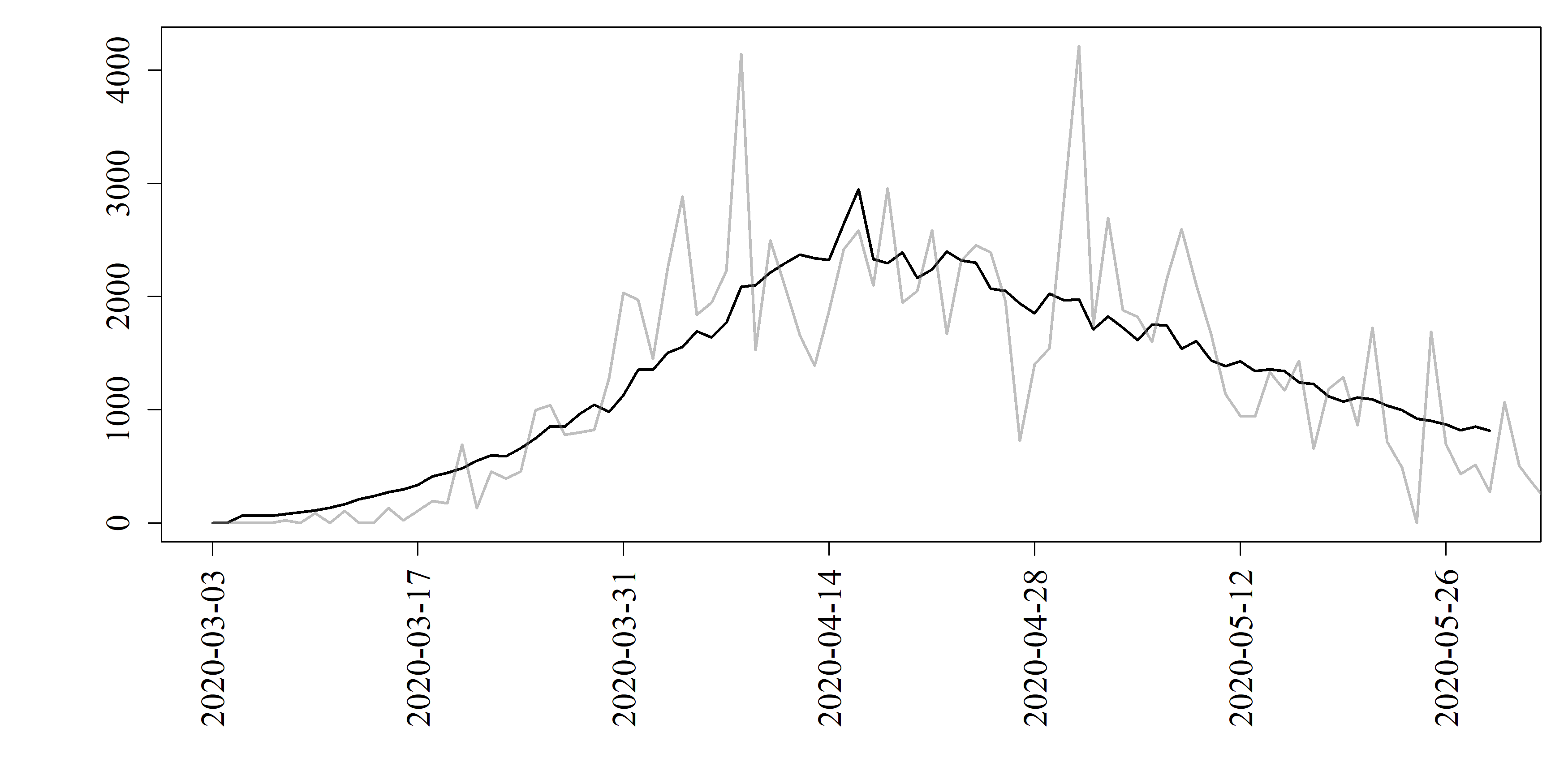}
    \caption{}
    \label{fig:gof-a}
    \end{subfigure}
    \begin{subfigure}[a]{\textwidth}
    \includegraphics[width=0.7\textwidth]{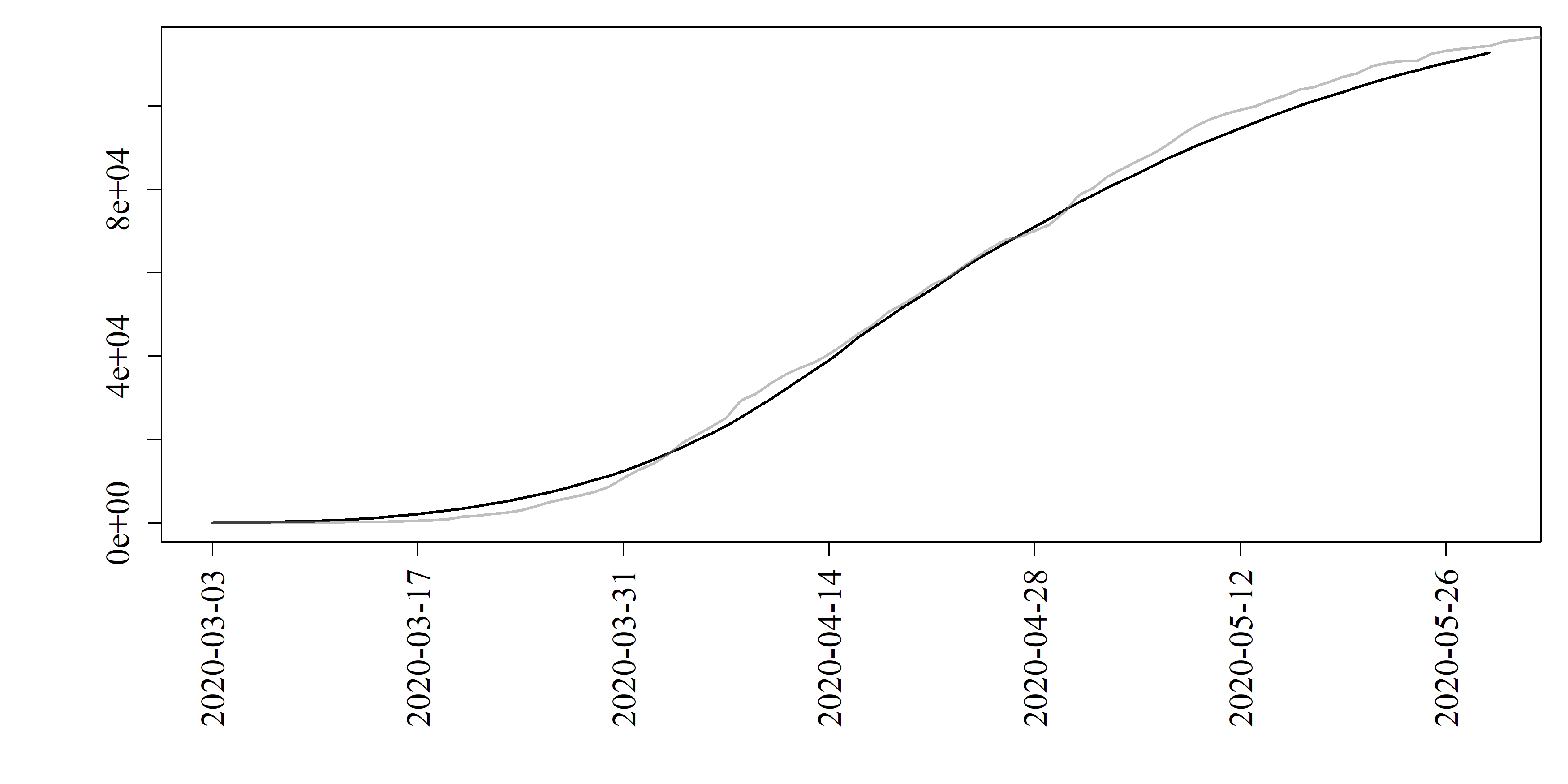}
    \caption{}
    \label{fig:gof-b}
    \end{subfigure}
    \caption{{\bf Estimated and observed number of daily infections.}\\
    Fitting the individual-based model (black solid) to the (a) daily and (b) cumulative number of confirmed infections (gray solid).}
    \label{fig:gof}
\end{figure}

By May 28, 2020, there were 453 deaths recorded in D.C. due to COVID-19 infections and 8,492 lab confirmed cases.  Figure \ref{fig:gof} shows the observed daily and cumulative case counts data alongside the estimated expected number of confirmed cases.

We estimated an initial $R_0$ of 1.54, an exposed period of 3 days, an infectious period of 6 days, and a daily probability of an individual importing COVID-19 from outside of the population of $8.6\times10^{-5}$.  The estimated effect of the face mask mandate was a reduction in transmissibility ($p_t$) of 23\%.

\begin{figure}
    \centering
    \captionsetup[subfigure]{justification=centering}
    \begin{subfigure}[a]{\textwidth}
    \centering
    \includegraphics[width=0.7\textwidth]{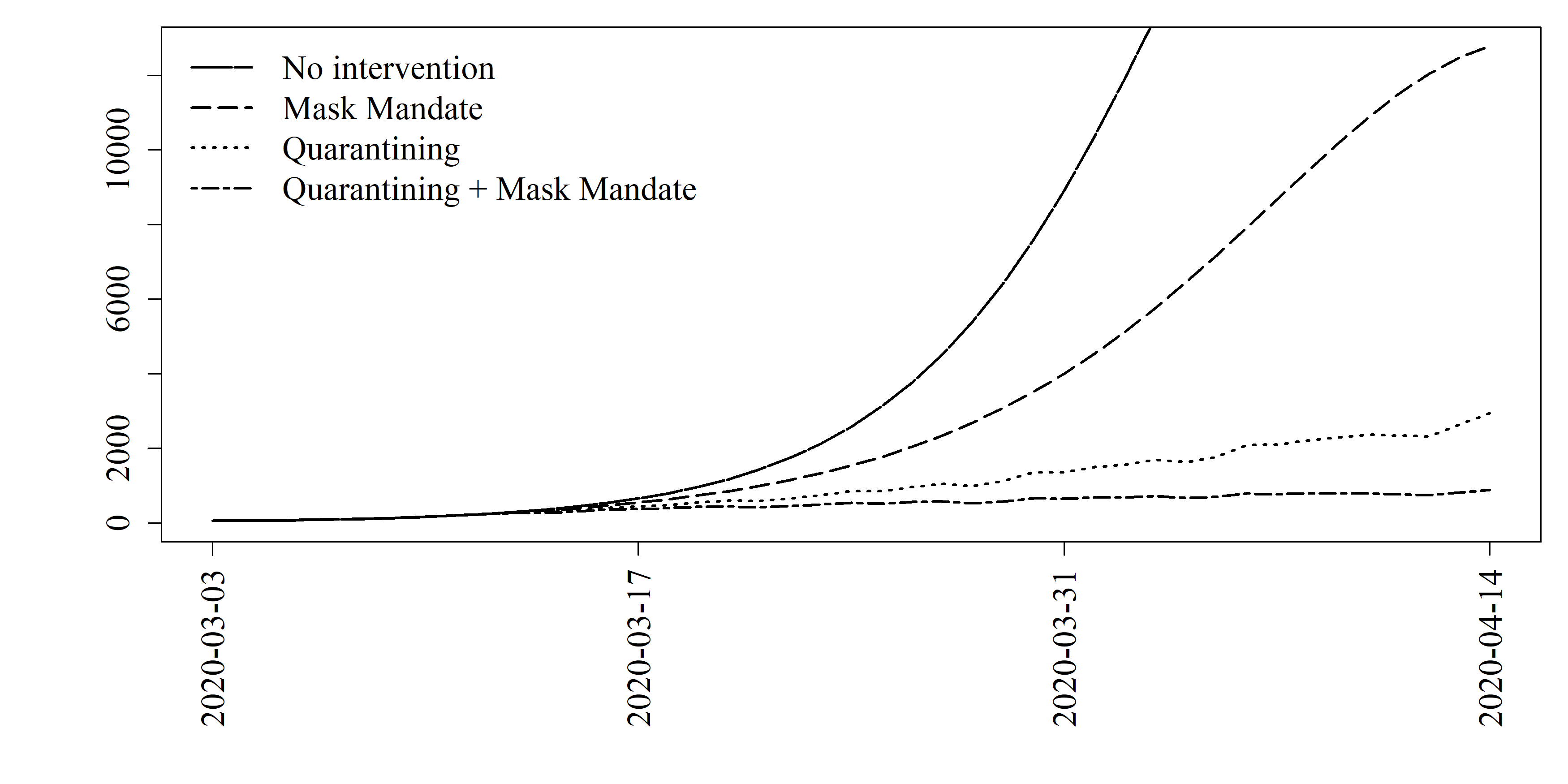}
    \caption{Interventions implemented on March 11,2020}
    \label{fig:results-a}
    \end{subfigure}\\
    \begin{subfigure}[a]{\textwidth}
    \centering
    \includegraphics[width=0.7\textwidth]{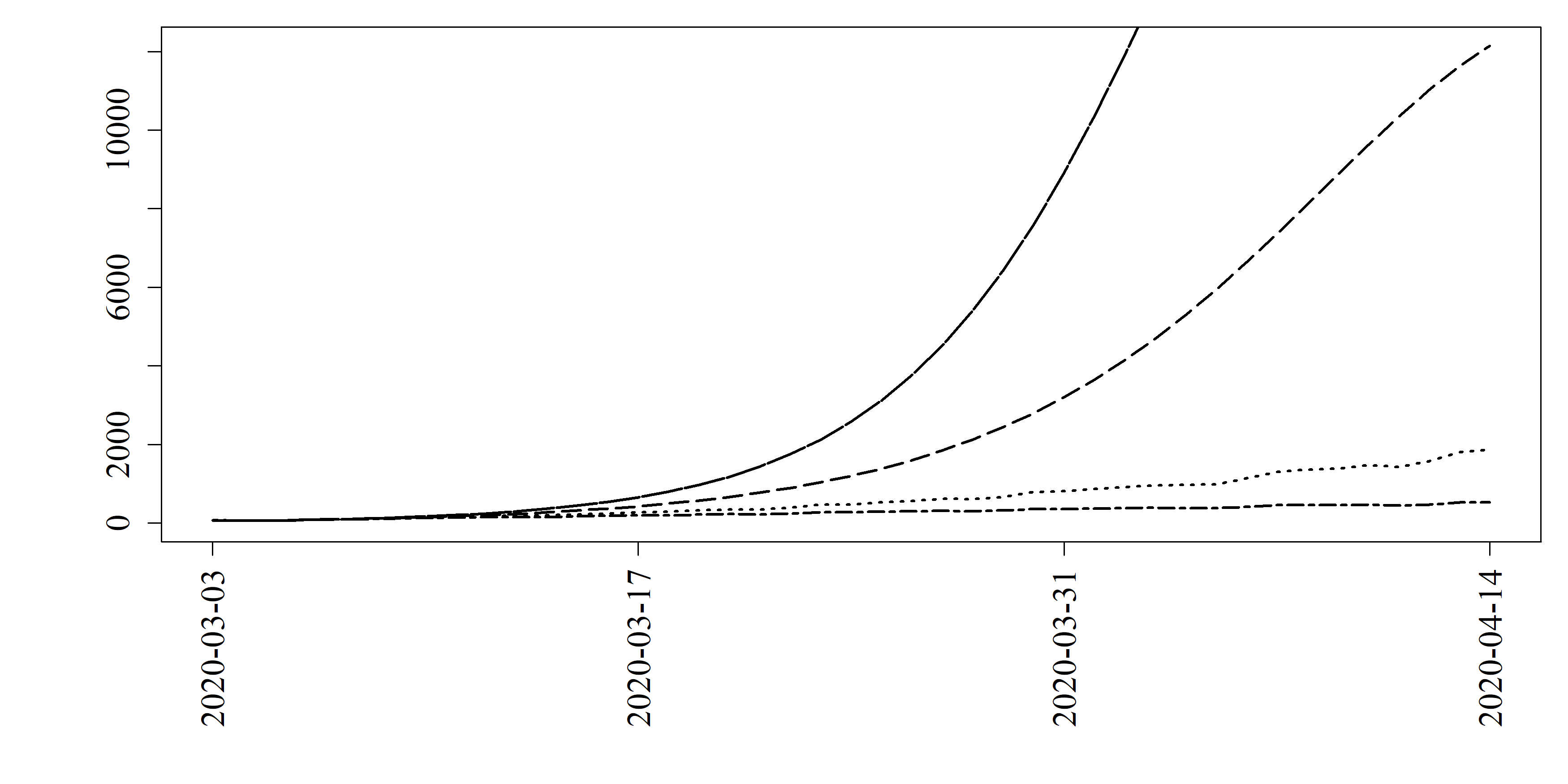}
    \caption{Interventions implemented on March 4,2020}
    \label{fig:results-b}
    \end{subfigure}\\
    \begin{subfigure}[a]{\textwidth}
    \centering
    \includegraphics[width=0.7\textwidth]{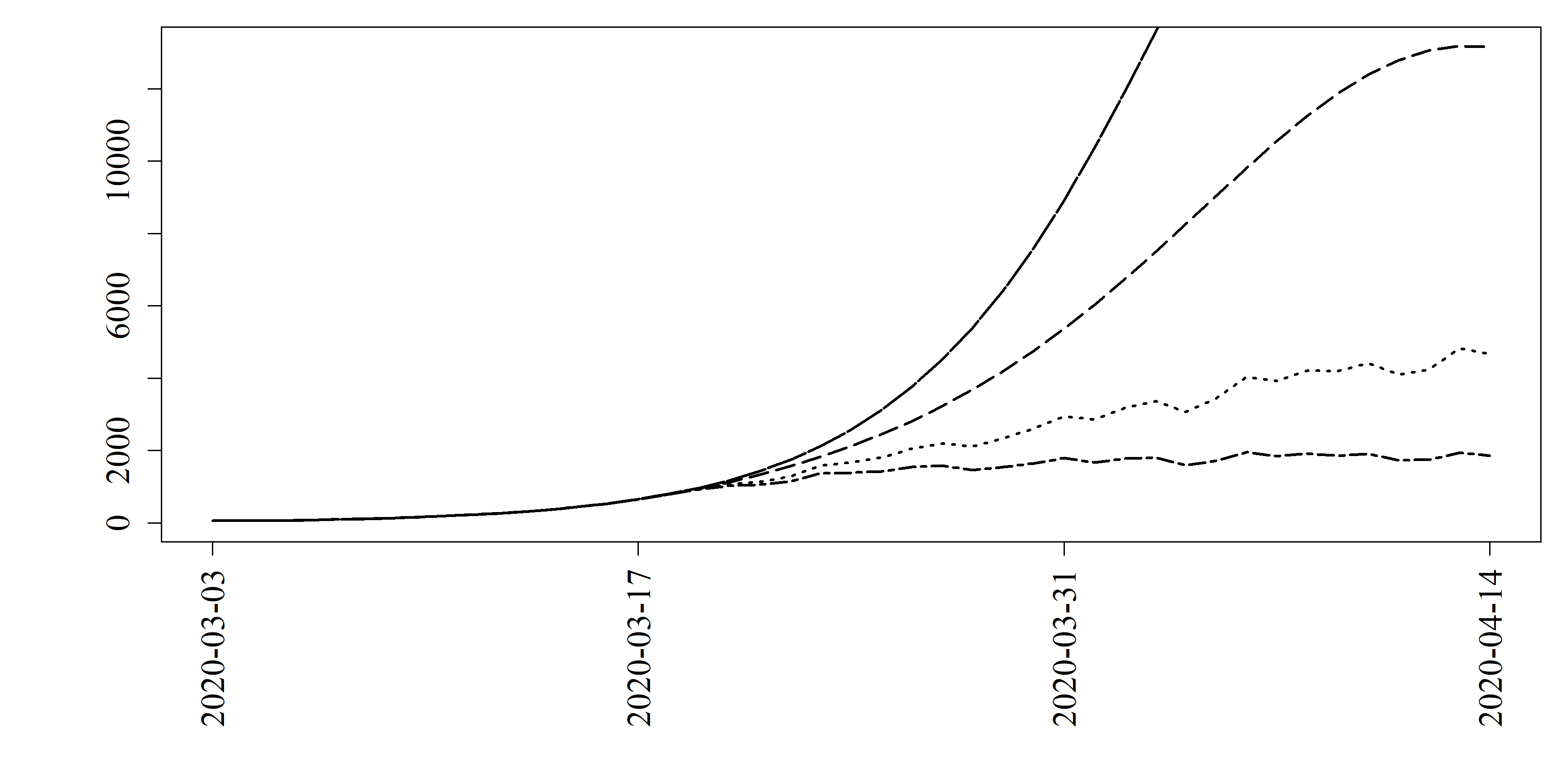}
    \caption{Interventions implemented on March 18,2020}
    \label{fig:results-c}
    \end{subfigure}
    \caption{{\bf Projections for nonpharmaceutical interventions.}\\
    Washington, D.C., COVID-19 projections for various mitigation strategies starting on, one week prior to, and one week after March 11, 2020. Quarantining is based on the changes in mobility due to the mayoral Declaration of Emergency. The face mask mandate matches the effects seen from the mayoral face mask mandate given on May 15, 2020 except with a one-week gradual rollout.}
    \label{fig:results}
\end{figure}

Figure \ref{fig:results} shows the main results for projecting under various mitigation measures through the date of the face mask mandate.  Each of the three subfigures corresponds to a given date of implementing the nonpharmaceutical interventions, March 11, March 4, and March 18, representing the actual date of intervention, the preceding week, and one extra week delay respectively. In each figure for comparison is the expected number of daily infections with no intervention (solid). Also given are curves representing the effects of a declaration of emergency (i.e., lockdown), a mask mandate, and both.  

\begin{figure}
    \centering
    \includegraphics[width=0.7\textwidth]{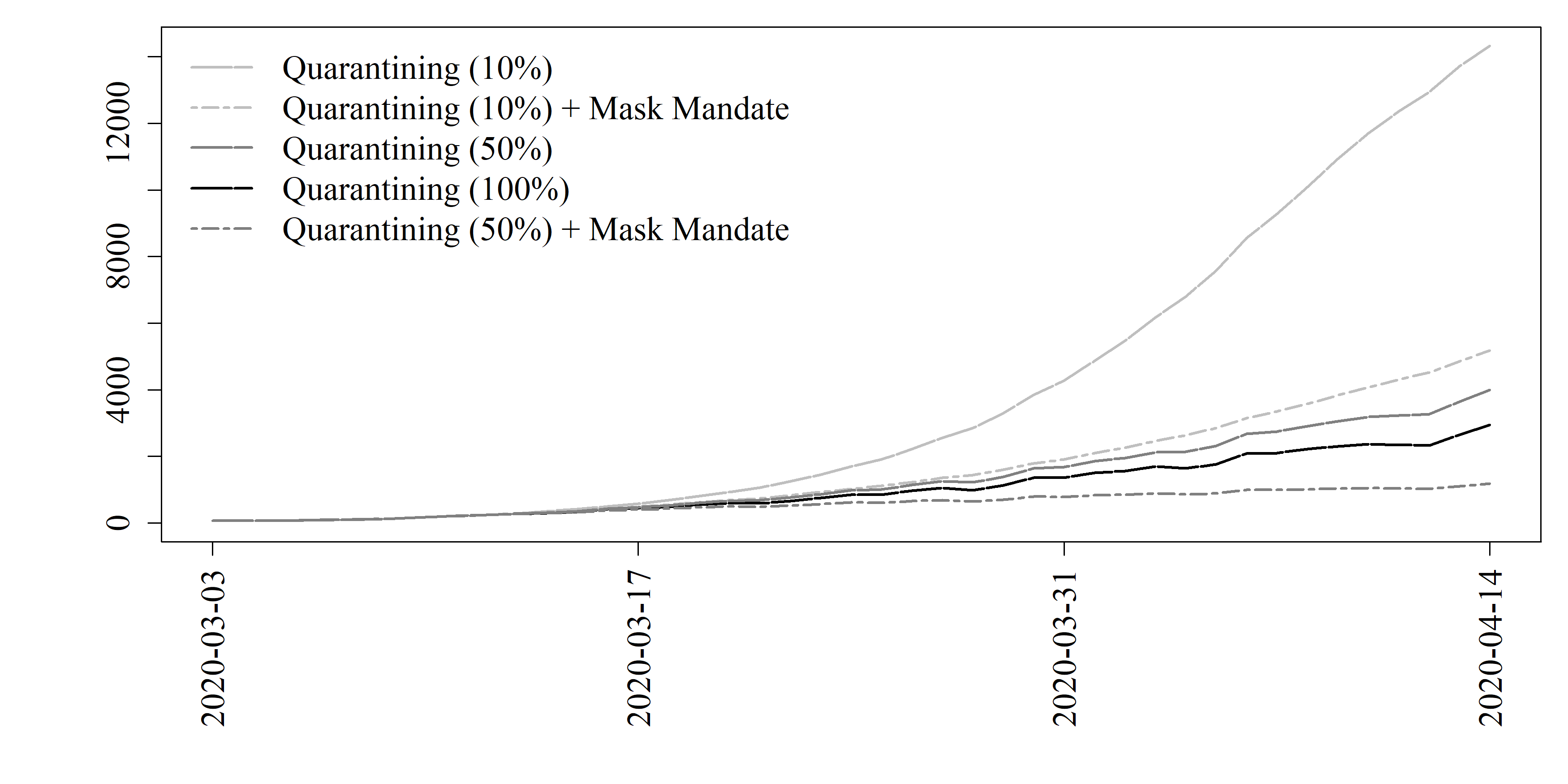}
    \caption{{\bf Projections for nonpharmaceutical interventions on subsets of the population.}\\
    Washington, D.C., COVID-19 projections for implementing quarantining for the top 10\% and 50\% individuals with respect to number of contacts.  Also shown are results for implementing quarantining for everyone.  Quarantining rates are the same observed levels regardless of the subpopulation practicing quarantining. The mask mandate applies to all individuals.}
    \label{fig:results-HD}
\end{figure}

Figure \ref{fig:results-HD} shows the results for contexts when only a subpopulation is expected to quarantine.  Quarantine rates, i.e., the daily probability that an individual will quarantine, are constant regardless of who is practicing quarantining.  Results are shown for the contexts where all are quarantining, and where the top 50\% and 10\% of the population in terms of number of contacts are practicing quarantining.  These strategies are also shown when augmented by a universal mask mandate applied to the entire population. 

\begin{figure}
    \centering
    \includegraphics[width=0.7\textwidth]{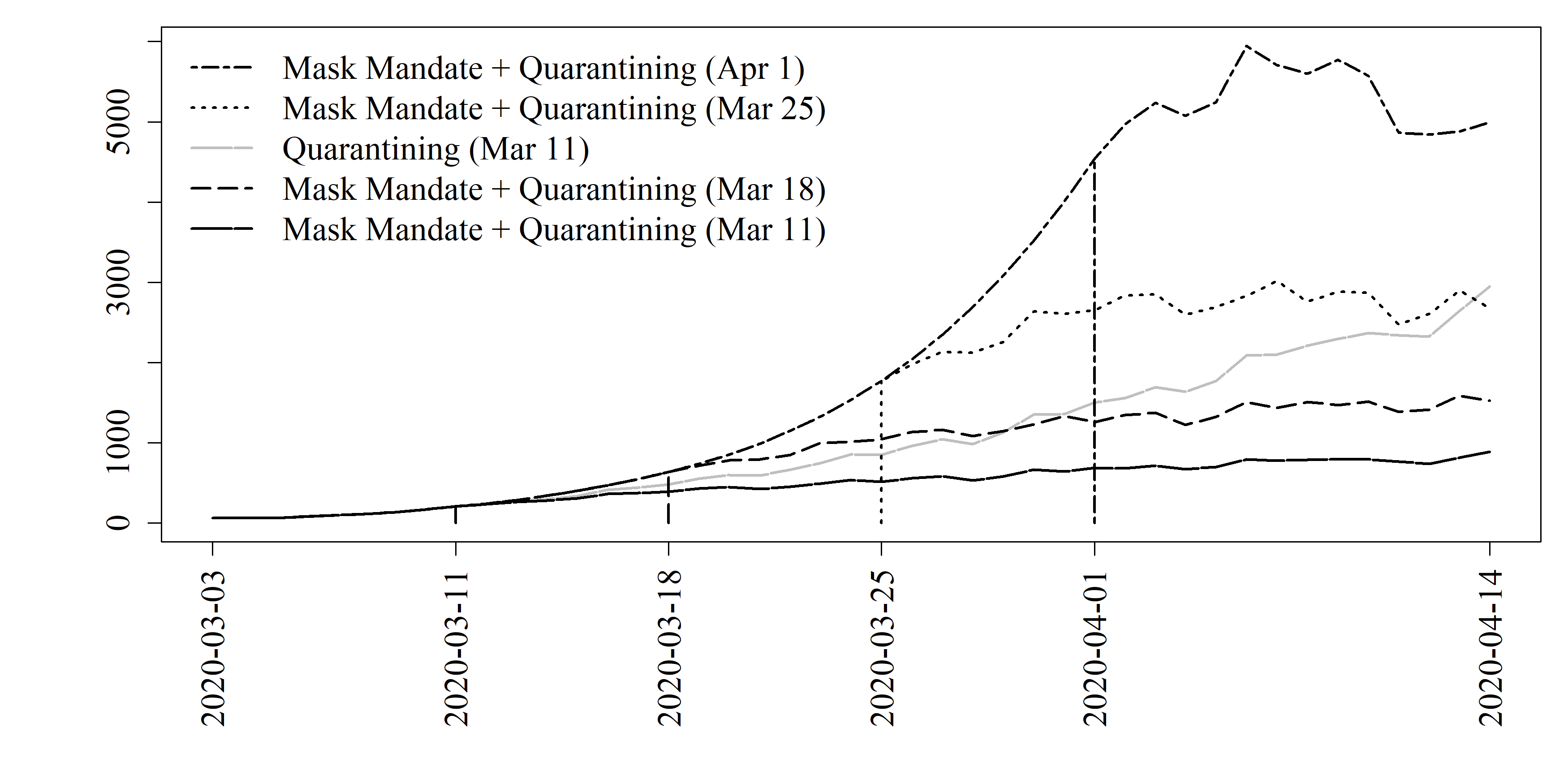}
    \caption{{\bf Projections for nonpharmaceutical interventions for a delayed lockdown.}\\
    Washington, D.C., COVID-19 projections for implementing the observed level of quarantining when quarantining is delayed by 1, 2, or 3 weeks.  This strategy is in conjunction with a mask mandate which begins on March 11 for all counterfactual scenarios. As a frame of reference, shown in gray is the result from a lockdown beginning on March 11 without a mask mandate, which corresponds to the actual policies enacted.}
    \label{fig:results-delayed}
\end{figure}

Figure \ref{fig:results-delayed} shows the results when the lockdown is delayed by 1, 2, or 3 weeks.  As a frame of reference, the curve which corresponds to the actual policies enacted (lockdown beginning on March 11 and no mask mandate).  Excepting this reference curve, all projections correspond to a mask mandate enacted on March 11 with a gradual rollout.  Quarantining begins on later  dates which are marked by the vertical line segments of matching line type.

\section*{Discussion}

In an ongoing outbreak, data streams are being continually updated, and estimates of the disease trajectory are highly dynamic.  Data and estimates quickly become outdated, and hence it is necessary to be able to provide accurate forecasts and projections of intervention effects rapidly.  In other contexts, researchers do not have access to a powerful computing cluster that allows one to perform the massive simulations required for an IBM.  Our approach provides an analytically tractable solution to compute the expected epidemic curve and is thus inherently fast and computationally inexpensive compared to simulation-based inference.

Our approach is limited in that in order to obtain credible or prediction intervals, simulation would still be required.  However, our work was motivated by the need to evaluate and compare a high volume of scenarios of implementing or ceasing nonpharmaceutical interventions, and our approach provides a fast and computationally inexpensive way to do this compared to simulation-based inference.

While we have focused on static networks, it would be a trivial extension of our work to account for time-varying networks.  Equation (\ref{eq:probH_it2}) can be changed to reflect this by replacing $A$ with some time-specific adjacency matrix  $A^{(t)}$, and all other formulas still hold. One important application of this would be to address the fact that some distancing measures affect age strata differently such as school or workplace closures.  By replacing the static network by a time-varying one, school closures, for example, could be reflected by having at a certain time $t$ the networks $\{A^{(s)}\}_{s\geq t}$ break contacts between most school-age children.  In addition, a vaccination pharmaceutical intervention applied at time $t$ could be addressed by breaking all contacts involving vaccinated individuals in the networks $\{A^{(s)}\}_{s\geq t}$.  As mentioned previously, our proposed approach cannot handle individual- or dyad-level varying transmission rates, and hence we cannot account for nonpharmaceutical or pharmaceutical interventions that act differentially in this way.

There are several key takeaways from our analysis of COVID-19 in D.C. First, an early lockdown augmented with a face mask mandate has a powerful effect on the expected epidemic trajectory. Adjusting for reporting rate, we estimate that there was a total of 44,590 infections by April 15.  Implementing a mask mandate alongside the lockdown on March 11, we estimate we would have expected to have seen a total of 20,616 infections.  Had these two interventions been implemented a week earlier we estimate we would have seen a total of 11,703 infections.  However, a second takeaway is that this effect is greatly diminished if delayed.  Delaying these two interventions by one week would have led to a total of 47,564.  Hence an early response was critical to mitigating the extent of COVID-19 spread.

Third, a mask mandate alone would probably not be sufficient.  Our results show that there would have been a dramatic increase in infections had a face mask mandate been put in place in lieu of a lockdown. This conclusion comes with a strong caveat, however.  Because individuals were wearing masks prior to the mask mandate, it is almost certain that we are underestimating the effect of a mask mandate. That is, the effects of universal mask wearing is underestimated due to the fact that this effect was already partially in place due to voluntary mask wearing, and so the effect of a mask mandate we have estimated is in fact the effect of mandating masks on those who would not voluntarily wear one.

Fourth, if careful consideration were to allow the strategic selection of individuals to quarantine, we could have seen almost the same level of reduction in infections by only implementing the lockdown for the top 50\% of individuals as ordered by number of contacts. This result could have important economic impacts, in that only half the population would need to quarantine.  We reiterate that this does not mean that half the population is always quarantined, but rather this half of the population is quarantining at the same rate as what we observed from the data for the overall population (see Figure \ref{fig:mobility}).  This strategic selection has a limit.  Our results demonstrate that only quarantining the top 10\% of individuals would lead to a much worse epidemic trajectory.  However, if this strategic 50\% quarantining strategy were paired with a mask mandate, we would have expected to see only 25,020 infections, just 56\% of the expected number of infections when quarantining all individuals without a mask mandate.

Fifth, by implementing a face mask mandate we can achieve similar or better results when delaying the lockdown by one to two weeks compared to an earlier lockdown without the mask mandate.  Given the heavy economic cost of the lockdown\cite{coibion2020cost}, this result has potentially strong consequences. Moreover, there is growing evidence of the clinical and public health related harms that may be attributable to prolonged lockdowns and societal quarantine; examples include delayed medical care \cite{lazzerini2020delayed,czeisler2020delay}, vaccination \cite{santoli2020effects}, cancer screening \cite{de2020cancer}, increased mental health problems \cite{brooks2020psychological}  and food insecurity \cite{loopstra2020vulnerability,hamadani2020immediate}. Thus, there are potentially many other indirect benefits to mask mandates that we do not account for in our model.

One important point is the sensitivity of our results to two factors.  First, we made all inference based on a single network generated as described above.  We checked and confirmed that there was negligible variation in our results based on using other networks generated according to the same mechanism.  Second, we checked how the accuracy of our results fared due to the approximation used in (\ref{eq:probH_it1}).  To do this, we compared the mean epicurve using our proposed method shown in Figure \ref{fig:gof-a} to a Monte Carlo estimate obtained by simulation.  We also compared the estimated curves in Figure \ref{fig:results-a} to Monte Carlo estimates.  There was an apparent problem of scale during the peak of the epidemic, but the shape and conclusions drawn from the curves were the same for those from our proposed approach and the Monte Carlo estimates. See the Supplementary Material for these and other sensitivity/accuracy results.

Another important limitation that is worth noting with any study involving the effects of public health policy is the indirect link between behavior changes and policy. It is often impossible to fully disentangle the endogeneity of behavior changes, the size of the epidemic, and the timing public policy, when the counterfactual cannot be observed. It is likely that mask wearing, social distancing and self quarantine are also driven by personal precautions and would increase in absence of public policy mandates. Indeed, there is evidence that many behavior changes that resulted in social distancing preceded policy changes \cite{cronin2020private}. Thus, our results should be interpreted based on the assumed connection to behavior (e.g., the baseline, \emph{no intervention}, case may not reflect actual behavior in absence of any policy).

\FloatBarrier

\end{document}